\documentclass[11pt]{article}
\usepackage{times}
\usepackage{sidecap}
\usepackage{geometry}
\usepackage{floatrow}
\usepackage{float}
\usepackage{hyperref}
\usepackage{graphics,graphicx}
\usepackage{subcaption}
\usepackage{epsf,wrapfig}

\usepackage{natbib,bm}

\usepackage{paralist}
\let\olditem\item
\renewenvironment{thebibliography}[1]{%
  \section*{\refname}
  \let\par\relax
  \renewcommand{\item}[1][]{\olditem[\textbullet]}%
  \inparaenum}{\endinparaenum}

\newcounter{aff}
\newcommand{\affil}[2]{%
    \par\noindent%
  \refstepcounter{aff}\label{#1}%
  \textsuperscript{\theaff} \textit{#2}\par}
\newcommand{\affmark}[1]{\textsuperscript{\ref{#1}}}

\DeclareGraphicsExtensions{.jpg,.pdf,.png,.eps}

\usepackage{caption}
\usepackage{xcolor}

\usepackage{tcolorbox}

\bibpunct{(}{)}{;}{a}{}{,}

\usepackage{soul}

\definecolor{darkcyan}{rgb}{0.0, 0.0, 0.7}

\definecolor{boxcolor}{RGB}{240,240,240} 
\definecolor{figboxcolor}{RGB}{193, 229, 255}
\definecolor{boxcolorgray}{RGB}{227, 227, 227}
\definecolor{darkcyan}{rgb}{0.0, 0.0, 0.7}

\sethlcolor{boxcolorgray}

\def\gtrsim{\lower 2pt \hbox{$\, \buildrel {\scriptstyle >}\over
{\scriptstyle \sim}\,$}}
\def\lesssim{\lower 2pt \hbox{$\, \buildrel {\scriptstyle <}\over
{\scriptstyle \sim}\,$}}

\def\approxlt{\lower.2em\hbox{$\buildrel < \over \sim$}}
\def\approxgt{\lower.2em\hbox{$\buildrel > \over \sim$}}

\def\be{\begin{equation}}
\def\ee{\end{equation}}
\def\bea{\begin{eqnarray*}}
\def\eea{\end{eqnarray*}}

\def\jref@jnl#1{{\rm#1}}
\def\aj{\jref@jnl{AJ}}			
\def\araa{\jref@jnl{ARA\&A}}		
\def\apj{\jref@jnl{ApJ}}			
\def\apjl{\jref@jnl{ApJ}}		
\def\apjs{\jref@jnl{ApJS}}		
\def\ao{\jref@jnl{Appl.~Opt.}}		
\def\apss{\jref@jnl{Ap\&SS}}		
\def\aap{\jref@jnl{A\&A}}		
\def\aapr{\jref@jnl{A\&A~Rev.}}		
\def\aaps{\jref@jnl{A\&AS}}		
\def\azh{\jref@jnl{AZh}}			
\def\baas{\jref@jnl{BAAS}}		
\def\jrasc{\jref@jnl{JRASC}}		
\def\memras{\jref@jnl{MmRAS}}		
\def\mnras{\jref@jnl{MNRAS}}		
\def\pra{\jref@jnl{Phys.~Rev.~A}}	
\def\prb{\jref@jnl{Phys.~Rev.~B}}	
\def\prc{\jref@jnl{Phys.~Rev.~C}}	
\def\prd{\jref@jnl{Phys.~Rev.~D}}	
\def\pre{\jref@jnl{Phys.~Rev.~E}}	
\def\prl{\jref@jnl{Phys.~Rev.~Lett.}}	
\def\pasp{\jref@jnl{PASP}}		
\def\pasj{\jref@jnl{PASJ}}		
\def\qjras{\jref@jnl{QJRAS}}		
\def\skytel{\jref@jnl{S\&T}}		
\def\solphys{\jref@jnl{Sol.~Phys.}}	
\def\sovast{\jref@jnl{Soviet~Ast.}}	
\def\ssr{\jref@jnl{Space~Sci.~Rev.}}	
\def\zap{\jref@jnl{ZAp}}			
\def\nat{\jref@jnl{Nature}}		
\def\iaucirc{\jref@jnl{IAU~Circ.}}
\def\aplett{\jref@jnl{Astrophys.~Lett.}}
\def\apspr{\jref@jnl{Astrophys.~Space~Phys.~Res.}}
\def\bain{\jref@jnl{Bull.~Astron.~Inst.~Netherlands}}
\def\fcp{\jref@jnl{Fund.~Cosmic~Phys.}}
\def\gca{\jref@jnl{Geochim.~Cosmochim.~Acta}}
\def\grl{\jref@jnl{Geophys.~Res.~Lett.}}
\def\jcp{\jref@jnl{J.~Chem.~Phys.}}	
\def\jgr{\jref@jnl{J.~Geophys.~Res.}}	
\def\jqsrt{\jref@jnl{J.~Quant.~Spec.~Radiat.~Transf.}}
\def\memsai{\jref@jnl{Mem.~Soc.~Astron.~Italiana}}
\def\nphysa{\jref@jnl{Nucl.~Phys.~A}}
\def\physrep{\jref@jnl{Phys.~Rep.}}
\def\physscr{\jref@jnl{Phys.~Scr}}
\def\planss{\jref@jnl{Planet.~Space~Sci.}}
\def\procspie{\jref@jnl{Proc.~SPIE}}

\def\ion#1#2{#1$\;${\small\rm\@Roman{#2}}\relax}

\newcount\lecurrentfam
\def\LaTeX{\lecurrentfam=\the\fam \leavevmode L\raise.42ex
\hbox{$\fam\lecurrentfam\scriptstyle\kern-.3em A$}\kern-.15em\TeX}
\def\sizrpt{
(\fontname\the\font): em=\the\fontdimen6\font, ex=\the\fontdimen5\font
\typeout{
(\fontname\the\font): em=\the\fontdimen6\font, ex=\the\fontdimen5\font
}}
\definecolor{dark}{rgb}{ 0.34, 0.09, 0.49}

\usepackage{fancyhdr,url}
\usepackage{transparent}
\usepackage{tablefootnote}
\usepackage{mathpazo}

\begin{document}

\noindent
{\Huge\bfseries
What Builds and Quenches the Most Massive Galaxies in the Early Universe?
\\[0.5em]
\Large 
A Wide-Field FIR/Sub-mm view with a new facility
}

\vfill
\thispagestyle{empty}
\noindent
\vspace{1em}

\vspace{1em}

\noindent Mengyuan Xiao\affmark{unige},
\noindent Longji Bing\affmark{Sussex},
\noindent Guilaine Lagache\affmark{LAM},
\noindent Miroslava Dessauges-Zavadsky\affmark{unige},
\noindent Olivier Ilbert \affmark{LAM},
\noindent Benjamin Magnelli \affmark{cea},
\noindent Pascal A. Oesch \affmark{unige}

\vspace{0.5em}

\affil{unige}{Department of Astronomy, University of Geneva, Chemin Pegasi 51, 1290 Versoix, Switzerland}

\affil{Sussex}{Astronomy Centre, University of Sussex, Falmer, Brighton BN1 9QH, UK} 

\affil{LAM}{Aix Marseille Univ, CNRS, CNES, LAM, Marseille, France} 

\affil{cea}{Université Paris-Saclay, Université Paris Cité, CEA, CNRS, AIM, 91191 Gif-sur-Yvette, France} 

\vspace{2em}

\footnotetext[1]{\url{mengyuan.xiao@unige.ch}}

\vspace{2em}
\vfill

\vfill
{
\begin{flushleft}
\end{flushleft}
}

\newpage{}

\begin{tcolorbox}[boxsep=3pt,left=0pt,right=0pt,top=1pt,bottom=1pt,colframe=boxcolor,colback=boxcolor]
\justify
\textbf{Abstract:} The first few billion years of cosmic history witnessed the rapid emergence of the most massive galaxies, yet their true space density, baryon assembly pathways, and early quenching mechanisms remain poorly constrained. Current surveys lack the wide-field, rest-frame FIR sensitivity needed to obtain a complete census of massive systems and to trace their cold gas, dust, and diffuse emission on galactic and circumgalactic scales. 
A next-generation facility with a very large aperture, wide field of view, and high mapping speed is essential to carry out deep, degree-scale rest-frame FIR surveys. Such capabilities are required to determine how common massive galaxies are, how they assemble their baryons, and what physical processes drive their early transformation and quenching.
\end{tcolorbox}

\vspace{-0.2cm}

\section{Scientific Context and Open Questions}
\vspace{-0.2cm}
In the standard cosmological framework, massive galaxies are expected to assemble gradually through gas accretion, star formation, and hierarchical mergers within evolving dark-matter halos. While this paradigm successfully reproduces many global trends across cosmic time, the first three years of the James Webb Space Telescope (JWST) have significantly changed our view of early galaxy assembly. JWST has revealed that the most massive galaxies formed bigger, faster, and earlier than anticipated: ultra-massive systems, already approaching Milky Way stellar mass, are present within the first billion years and include dust-obscured systems, compact quiescent galaxies, and morphologically mature disks with bulge components \citep[e.g., ][]{Boylan-Kolchin2023,Dekel2023, Akins2023, Xiao2024, Weibel2024, Carnall2023Natur, glazebrook2017, deGraaff2024, Xiao2025_spiral}. 
These discoveries raise fundamental questions about the abundance, baryon assembly histories, and transformation pathways of the most massive galaxies at early epochs. Yet our current observational view remains incomplete.

\begin{figure}[!tbh]
    \centering
    \includegraphics[width=1\textwidth]{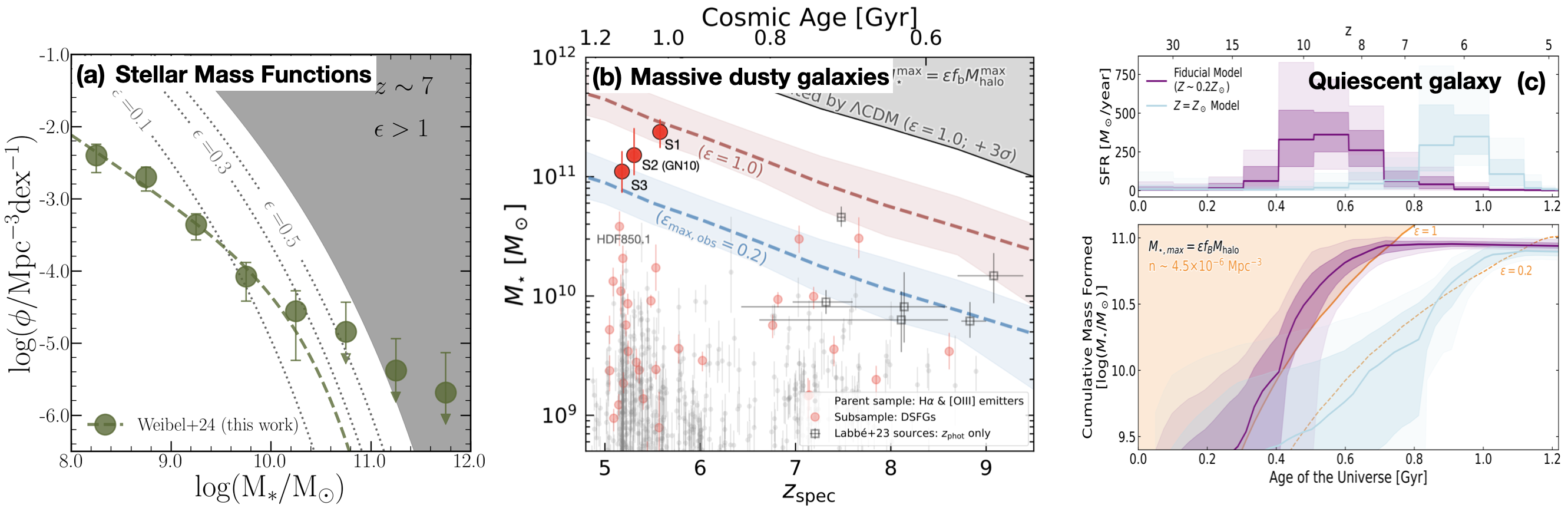}
    \vspace{-8mm}
    \caption{JWST indicates the efficient formation of massive galaxies. \textbf{a}: large excess at the high mass end of the stellar mass function \citep{Weibel2024}; \textbf{b}: three massive dusty galaxies at $z_{spec}=5-6$ showing efficient star formation -- about 50\% on average of the baryons in their halos to be converted into stars \citep{Xiao2024}; \textbf{c}: the cumulative mass assembly history of a massive quiescent galaxy $z_{spec} = 4.8963$ implies a high efficiency of star formation by converting $>$20\% baryons into stars \citep{deGraaff2024}.}
    \label{fig:1}
\end{figure}

Two limitations dominate present surveys.  
First, the deepest extragalactic fields span only $\sim$0.01–0.5 deg$^{2}$ \citep[e.g.,][]{Oesch2023, Casey2023, Gomez-Guijarro2022a}, leading to severe cosmic variance and undersampling of the rare, massive halos where the most massive galaxies reside.  
Second, we lack wide-field sensitivity in the rest-frame far-infrared (FIR), where dust reprocesses the majority of stellar radiation and where key cooling lines trace the cold interstellar and circumgalactic media.  
JWST offers unprecedented sensitivity in the rest-frame optical and near-infrared, but cannot probe the FIR, and even wide-area optical/near-IR missions such as Euclid will map large sky areas without providing FIR coverage \citep[e.g.][]{EuclidOverview}.     
Interferometric sub-mm facilities (e.g., ALMA, NOEMA) provide exquisite resolution but can neither survey large areas at the necessary depth nor recover low-surface-brightness diffuse emission. 
As a result, current datasets do not provide a complete census of massive galaxies or the FIR diagnostics necessary to understand their growth and transformation.

A next-generation, wide-field, and sensitive FIR facility is therefore essential to:  
(1) obtain a complete, unbiased census of massive galaxies,  
(2) map the baryonic reservoirs that fuel their rapid early growth, and  
(3) identify the physical processes responsible for their early transition from star-forming to quiescence.  
A future single-dish telescope (with $\sim$50-m aperture allowing for a spatial resolution of $\sim$2'' at 600\,GHz), large field of view, high mapping speed, and high sensitivity to all spatial scales, is uniquely suited to address these challenges.
\par\smallskip\noindent
Three overarching scientific questions define the motivation for this white paper:
\begin{enumerate}
    \item \textbf{What is the true space density of the most massive galaxies in the first few billion years?}  
    Current surveys sample insufficient cosmological volume and miss heavily obscured systems, leaving the abundance and distribution of early massive galaxies poorly constrained.

    \item \textbf{How do the most massive galaxies assemble their baryons across the first few Gyr?}  
    We lack statistical measurements of cold gas, dust, and metals across representative massive galaxy populations and their surrounding environments.

    \item \textbf{What physical processes drive the early transformation and quenching of massive galaxies?}  
    The mechanisms behind rapid quenching, including gas exhaustion, feedback, and environmental effects, remain uncertain and require FIR constraints on gas content, depletion timescales, and the large-scale structures they reside in.
\end{enumerate}

\vspace{-0.5cm}

\section{Science Case}
\vspace{-0.2cm}
\subsection{The Space Density of Massive Galaxies in the Early Universe}
\vspace{-0.2cm}
A robust determination of the abundance of massive galaxies requires deep, wide-field FIR surveys capable of detecting both normal and heavily dust-obscured systems. 
Rest-frame optical/near-infrared wide-area surveys inherently miss obscured galaxies \citep[e.g., HST-dark, and even JWST NIRCam-dark galaxies;][]{wangtao2019, franco2018, Xiao2023, Barrufet2025, Bing2025}, while targeted sub-mm follow-up cannot deliver statistically representative samples. 
Besides, present blind surveys either originate primarily from small-area fields highly susceptible to cosmic variance and lacking rare objects hosted by massive halos.
A new facility is needed to enable degree-scale surveys in rest-frame FIR with the depth required to obtain a complete census of massive galaxies across cosmologically representative volumes. 

These measurements are essential for constraining the high-mass end of the stellar mass function and IR luminosity function jointly with multiwavelength data, testing predictions of halo growth, and anchoring models of early structure formation.

\vspace{-0.2cm}
\subsection{Baryon Assembly in Massive Galaxies}
\vspace{-0.2cm}
Understanding how massive galaxies acquire, retain, and process their baryonic content requires simultaneous constraints on cold gas, dust, and metals.  
FIR continuum emission traces dust mass and obscured star formation, while bright FIR fine-structure lines (e.g., [C\,{\sc ii}], [O\,{\sc iii}]) probe the multiphase ISM and metal enrichment and connect galaxies to their circumgalactic environments \citep[e.g.,][]{Ginolfi2020,Lagache2025}.  
Existing facilities cannot provide the combination of:  
(1) wide-area coverage,  
(2) sensitivity to both compact and diffuse emission, and  
(3) access to the spatial scales needed to connect galaxies to their larger environment.  
We need a facility to measure gas fractions, star formation efficiency, metal enrichment, and environmental dependence across thousands of massive galaxies to provide the first coherent picture of early baryon assembly.

\subsection{Early Transformation and Quenching of Massive Galaxies}
\vspace{-0.2cm}
Massive quiescent galaxies are found to emerge just within the first few billion years \citep[e.g.][]{Valentino2023}, yet the physical mechanisms responsible for their quenching remain unclear.  
The lack of gas measurements for representative quiescent and transitioning galaxies, the difficulty of detecting multiphase gas associated with feedback or environmental processes, and the absence of a wide-field environmental context all limit our ability to test different quenching scenarios \citep[e.g.][]{Martig2009,Harrison2018}.  
A $\sim$50-m aperture single-dish telescope can uniquely constrain cold gas and dust in individual massive quenching galaxies with a gas fraction of a few percent \citep{D'Eugenio2023}. With the large field of view, it could also simultaneously probe the large-scale structures and environmental effects through the gas and dust in companion galaxies and the thermal Sunyaev-Zeldovich signal from the surrounding heated gas.  
By combining wide-field FIR mapping with multi-wavelength data from JWST, ALMA, SKA and ELTs, 
such a facility will help distinguish quenching pathways, measure depletion timescales, and probe the connection between quiescent galaxies and the large-scale environments they reside in.

\vspace{-0.2cm}
\section{Technical Requirements}
\vspace{-0.2cm}
To probe the formation of early massive galaxies, we thus need:\\
- The coverage of (sub-)millimeter wavelength (from $\sim$ 30 to 950\,GHz) to probe the dust continuum emission and the redshifted lines (as [CII] and CO). The facility has thus to be constructed in a very dry environment (as that offered at 5000 m altitude near the ALMA site).\\
- A high angular resolution (diffraction-limited $\sim2''$ in the sub-millimeter), to beat the confusion noise and precisely locate the galaxies, thus requiring a telescope diameter of $\sim$ 50 meters.\\
- A high throughput, offered by the combination of a large field of view (1-2$^\circ$) and large aperture, to map large areas of the sky. Combined with a high sensitivity, such a facility has to offer fast mapping speed ($>10^3$ and up to 10$^5$ times better than ALMA).\\
- Wide-band on-chip spectrometers entirely cover one or a few atmospheric windows with R$\gtrsim$500 for spectral line detection, and multi-wavelength cameras observe the dust continuum, both filling the entire or a significant fraction of the field of view.

\par\medskip\noindent
\noindent
\bibliographystyle{apj_mod}
\bibliography{reference_update}

\begin{thebibliography}{24}
\expandafter\ifx\csname natexlab\endcsname\relax\def\natexlab#1{#1}\fi

\bibitem[{{Akins} {et~al.}(2023){Akins}, {Casey}, {Allen}, {Bagley},
  {Dickinson}, {Finkelstein}, {Franco}, {Harish}, {Arrabal Haro}, {Ilbert},
  {Kartaltepe}, {Koekemoer}, {Liu}, {Long}, {McCracken}, {Paquereau},
  {Papovich}, {Pirzkal}, {Rhodes}, {Robertson}, {Shuntov}, {Toft}, {Yang},
  {Barro}, {Bisigello}, {Buat}, {Champagne}, {Cooper}, {Costantin}, {de La
  Vega}, {Drakos}, {Faisst}, {Fontana}, {Fujimoto}, {Gillman},
  {G{\'o}mez-Guijarro}, {Gozaliasl}, {Hathi}, {Hayward}, {Hirschmann},
  {Holwerda}, {Jin}, {Kocevski}, {Kokorev}, {Lambrides}, {Lucas}, {Magdis},
  {Magnelli}, {McKinney}, {Mobasher}, {P{\'e}rez-Gonz{\'a}lez}, {Rich},
  {Seill{\'e}}, {Talia}, {Urry}, {Valentino}, {Whitaker}, {Yung}, {Zavala},
  {Cosmos-Web Team}, \& {Ceers Team}}]{Akins2023}
\textbf{Akins} {et~al.} 2023, \apj, 956, 61.
\bibitem[{{Barrufet} {et~al.}(2025){Barrufet}, {Oesch}, {Marques-Chaves},
  {Arellano-Cordova}, {Baggen}, {Carnall}, {Cullen}, {Dunlop}, {Gottumukkala},
  {Fudamoto}, {Illingworth}, {Magee}, {McLure}, {McLeod}, {Micha{\l}owski},
  {Stefanon}, {van Dokkum}, \& {Weibel}}]{Barrufet2025}
\textbf{Barrufet} {et~al.} 2025, \mnras.
\bibitem[{{Bing} {et~al.}(2025){Bing}, {Oliver}, {Xiao}, {Lagache}, {Adscheid},
  {Liu}, {Magnelli}, {Neri}, {Dessauges-Zavadsky}, {Koekemoer}, {Franco},
  {Jin}, {Cooper}, {Faisst}, {Casey}, {Kartaltepe}, {Akins}, {Beelen}, {Elbaz},
  {Gillman}, {Harish}, {Long}, {McCracken}, {Oesch}, {Paquereau}, {Ponthieu},
  {Rhodes}, {Robertson}, {Sanders}, {Shuntov}, \& {Wilkins}}]{Bing2025}
\textbf{Bing} {et~al.} 2025, arXiv e-prints, arXiv:2511.08672.
\bibitem[{{Boylan-Kolchin}(2023)}]{Boylan-Kolchin2023}
\textbf{Boylan-Kolchin}, M. 2023, Nature Astronomy, 7, 731.
\bibitem[{{Carnall} {et~al.}(2023){Carnall}, {McLure}, {Dunlop}, {McLeod},
  {Wild}, {Cullen}, {Magee}, {Begley}, {Cimatti}, {Donnan}, {Hamadouche},
  {Jewell}, \& {Walker}}]{Carnall2023Natur}
\textbf{Carnall} {et~al.} 2023, \nat, 619, 716.
\bibitem[{{Casey} {et~al.}(2023){Casey}, {Kartaltepe}, {Drakos}, {Franco},
  {Harish}, {Paquereau}, {Ilbert}, {Rose}, {Cox}, {Nightingale}, {Robertson},
  {Silverman}, {Koekemoer}, {Massey}, {McCracken}, {Rhodes}, {Akins}, {Allen},
  {Amvrosiadis}, {Arango-Toro}, {Bagley}, {Bongiorno}, {Capak}, {Champagne},
  {Chartab}, {Ch{\'a}vez Ortiz}, {Chworowsky}, {Cooke}, {Cooper}, {Darvish},
  {Ding}, {Faisst}, {Finkelstein}, {Fujimoto}, {Gentile}, {Gillman}, {Gould},
  {Gozaliasl}, {Hayward}, {He}, {Hemmati}, {Hirschmann}, {Jahnke}, {Jin},
  {Khostovan}, {Kokorev}, {Lambrides}, {Laigle}, {Larson}, {Leung}, {Liu},
  {Liaudat}, {Long}, {Magdis}, {Mahler}, {Mainieri}, {Manning}, {Maraston},
  {Martin}, {McCleary}, {McKinney}, {McPartland}, {Mobasher}, {Pattnaik},
  {Renzini}, {Rich}, {Sanders}, {Sattari}, {Scognamiglio}, {Scoville}, {Sheth},
  {Shuntov}, {Sparre}, {Suzuki}, {Talia}, {Toft}, {Trakhtenbrot}, {Urry},
  {Valentino}, {Vanderhoof}, {Vardoulaki}, {Weaver}, {Whitaker}, {Wilkins},
  {Yang}, \& {Zavala}}]{Casey2023}
\textbf{Casey} {et~al.} 2023, \apj, 954, 31.
\bibitem[{{de Graaff} {et~al.}(2025){de Graaff}, {Setton}, {Brammer}, {Cutler},
  {Suess}, {Labb{\'e}}, {Leja}, {Weibel}, {Maseda}, {Whitaker}, {Bezanson},
  {Boogaard}, {Cleri}, {De Lucia}, {Franx}, {Greene}, {Hirschmann}, {Matthee},
  {McConachie}, {Naidu}, {Oesch}, {Price}, {Rix}, {Valentino}, {Wang}, \&
  {Williams}}]{deGraaff2024}
\textbf{de Graaff} {et~al.} 2025, Nature Astronomy, 9, 280.
\bibitem[{{Dekel} {et~al.}(2023){Dekel}, {Sarkar}, {Birnboim}, {Mandelker}, \&
  {Li}}]{Dekel2023}
\textbf{Dekel} {et~al.} 2023, \mnras, 523, 3201.
\bibitem[{{D'Eugenio} {et~al.}(2023){D'Eugenio}, {Daddi}, {Liu}, \&
  {Gobat}}]{D'Eugenio2023}
\textbf{D'Eugenio} {et~al.} 2023, \aap, 678, L9.
\bibitem[{{Euclid Collaboration} {et~al.}(2025){Euclid Collaboration},
  {Mellier}, {Abdurro'uf}, {Acevedo Barroso}, {Ach{\'u}carro}, {Adamek},
  {Adam}, {Addison}, {Aghanim}, {Aguena}, {Ajani}, {Akrami}, {Al-Bahlawan},
  {Alavi}, {Albuquerque}, {Alestas}, {Alguero}, {Allaoui}, {Allen}, {Allevato},
  {Alonso-Tetilla}, {Altieri}, {Alvarez-Candal}, {Alvi}, {Amara}, {Amendola},
  {Amiaux}, {Andika}, {Andreon}, {Andrews}, {Angora}, {Angulo}, {Annibali},
  {Anselmi}, {Anselmi}, {Arcari}, {Archidiacono}, {Aric{\`o}}, {Arnaud},
  {Arnouts}, {Asgari}, {Asorey}, {Atayde}, {Atek}, {Atrio-Barandela}, {Aubert},
  {Aubourg}, {Auphan}, {Auricchio}, {Aussel}, {Aussel}, {Avelino},
  {Avgoustidis}, {Avila}, {Awan}, {Azzollini}, {Baccigalupi}, {Bachelet},
  {Bacon}, {Baes}, {Bagley}, {Bahr-Kalus}, {Balaguera-Antolinez}, {Balbinot},
  {Balcells}, {Baldi}, {Baldry}, {Balestra}, {Ballardini}, {Ballester},
  {Balogh}, {Ba{\~n}ados}, {Barbier}, {Bardelli}, {Baron}, {Barreiro},
  {Barrena}, {Barriere}, {Barros}, {Barthelemy}, {Bartolo}, {Basset},
  {Battaglia}, {Battisti}, {Baugh}, {Baumont}, {Bazzanini}, {Beaulieu},
  {Beckmann}, {Belikov}, {Bel}, {Bellagamba}, {Bella}, {Bellini}, {Benabed},
  {Bender}, {Benevento}, {Bennett}, {Benson}, {Bergamini}, {Bermejo-Climent},
  {Bernardeau}, {Bertacca}, {Berthe}, {Berthier}, {Bethermin}, {Beutler},
  {Bevillon}, {Bhargava}, {Bhatawdekar}, {Bianchi}, {Bisigello}, {Biviano},
  {Blake}, {Blanchard}, {Blazek}, {Blot}, {Bosco}, {Bodendorf}, {Boenke},
  {B{\"o}hringer}, {Boldrini}, {Bolzonella}, {Bonchi}, {Bonici}, {Bonino},
  {Bonino}, {Bonvin}, {Bon}, {Booth}, {Borgani}, {Borlaff}, {Borsato}, {Bose},
  {Botticella}, {Boucaud}, {Bouche}, {Boucher}, {Boutigny}, {Bouvard},
  {Bouwens}, {Bouy}, {Bowler}, {Bozza}, {Bozzo}, {Branchini}, {Brando},
  {Brau-Nogue}, {Brekke}, {Bremer}, {Brescia}, {Breton}, {Brinchmann},
  {Brinckmann}, {Brockley-Blatt}, {Brodwin}, {Brouard}, {Brown}, {Bruton},
  {Bucko}, {Buddelmeijer}, {Buenadicha}, {Buitrago}, {Burger}, {Burigana},
  {Busillo}, {Busonero}, {Cabanac}, {Cabayol-Garcia}, {Cagliari}, {Caillat},
  {Caillat}, {Calabrese}, {Calabro}, {Calderone}, {Calura}, {Camacho Quevedo},
  {Camera}, {Campos}, {Ca{\~n}as-Herrera}, {Candini}, {Cantiello},
  {Capobianco}, {Cappellaro}, {Cappelluti}, {Cappi}, {Caputi}, {Cara},
  {Carbone}, {Cardone}, {Carella}, {Carlberg}, {Carle}, {Carminati}, {Caro},
  {Carrasco}, {Carretero}, {Carrilho}, {Carron Duque}, \&
  {Carry}}]{EuclidOverview}
\textbf{Euclid Collaboration} {et~al.} 2025, \aap, 697, A1.
\bibitem[{{Franco} {et~al.}(2018){Franco}, {Elbaz}, {B{\'e}thermin},
  {Magnelli}, {Schreiber}, {Ciesla}, {Dickinson}, {Nagar}, {Silverman},
  {Daddi}, {Alexander}, {Wang}, {Pannella}, {Le Floc'h}, {Pope}, {Giavalisco},
  {Maury}, {Bournaud}, {Chary}, {Demarco}, {Ferguson}, {Finkelstein}, {Inami},
  {Iono}, {Juneau}, {Lagache}, {Leiton}, {Lin}, {Magdis}, {Messias},
  {Motohara}, {Mullaney}, {Okumura}, {Papovich}, {Pforr}, {Rujopakarn},
  {Sargent}, {Shu}, \& {Zhou}}]{franco2018}
\textbf{Franco} {et~al.} 2018, \aap, 620, A152.
\bibitem[{{Ginolfi} {et~al.}(2020){Ginolfi}, {Jones}, {B{\'e}thermin},
  {Fudamoto}, {Loiacono}, {Fujimoto}, {Le F{\'e}vre}, {Faisst}, {Schaerer},
  {Cassata}, {Silverman}, {Yan}, {Capak}, {Bardelli}, {Boquien}, {Carraro},
  {Dessauges-Zavadsky}, {Giavalisco}, {Gruppioni}, {Ibar}, {Khusanova},
  {Lemaux}, {Maiolino}, {Narayanan}, {Oesch}, {Pozzi}, {Rodighiero}, {Talia},
  {Toft}, {Vallini}, {Vergani}, \& {Zamorani}}]{Ginolfi2020}
\textbf{Ginolfi} {et~al.} 2020, \aap, 633, A90.
\bibitem[{{Glazebrook} {et~al.}(2017){Glazebrook}, {Schreiber}, {Labb{\'e}},
  {Nanayakkara}, {Kacprzak}, {Oesch}, {Papovich}, {Spitler}, {Straatman},
  {Tran}, \& {Yuan}}]{glazebrook2017}
\textbf{Glazebrook} {et~al.} 2017, \nat, 544, 71.
\bibitem[{{G{\'o}mez-Guijarro} {et~al.}(2022){G{\'o}mez-Guijarro}, {Elbaz},
  {Xiao}, {B{\'e}thermin}, {Franco}, {Magnelli}, {Daddi}, {Dickinson},
  {Demarco}, {Inami}, {Rujopakarn}, {Magdis}, {Shu}, {Chary}, {Zhou},
  {Alexander}, {Bournaud}, {Ciesla}, {Ferguson}, {Finkelstein}, {Giavalisco},
  {Iono}, {Juneau}, {Kartaltepe}, {Lagache}, {Le Floc'h}, {Leiton}, {Lin},
  {Motohara}, {Mullaney}, {Okumura}, {Pannella}, {Papovich}, {Pope}, {Sargent},
  {Silverman}, {Treister}, \& {Wang}}]{Gomez-Guijarro2022a}
\textbf{G{\'o}mez-Guijarro} {et~al.} 2022, \aap, 658, A43.
\bibitem[{{Harrison} {et~al.}(2018){Harrison}, {Costa}, {Tadhunter},
  {Fl{\"u}tsch}, {Kakkad}, {Perna}, \& {Vietri}}]{Harrison2018}
\textbf{Harrison} {et~al.} 2018, Nature Astronomy, 2, 198.
\bibitem[{{Lagache} {et~al.}(2025){Lagache}, {Xiao}, {Beelen}, {Berta},
  {Ciesla}, {Neri}, {Pello}, {Adam}, {Ade}, {Ajeddig}, {Amarantidis},
  {Andr{\'e}}, {Aussel}, {Beno{\^\i}t}, {B{\'e}thermin}, {Bing}, {Bongiovanni},
  {Bounmy}, {Bourrion}, {Calvo}, {Catalano}, {Ch{\'e}rouvrier}, {Chowdhury},
  {De Petris}, {D{\'e}sert}, {Doyle}, {Driessen}, {Ejlali}, {Ferragamo},
  {Gomez}, {Goupy}, {Hanser}, {Katsioli}, {K{\'e}ruzor{\'e}}, {Kramer},
  {Ladjelate}, {Leclercq}, {Lestrade}, {Mac{\'\i}as-P{\'e}rez}, {Madden},
  {Maury}, {Mayet}, {Monfardini}, {Moyer-Anin}, {Mu{\~n}oz-Echeverr{\'\i}a},
  {Myserlis}, {Oesch}, {Paliwal}, {Perotto}, {Pisano}, {Ponthieu},
  {Rev{\'e}ret}, {Rigby}, {Ritacco}, {Roussel}, {Ruppin}, {S{\'a}nchez-Portal},
  {Savorgnano}, {Schuster}, {Sievers}, {Tucker}, \& {Zylka}}]{Lagache2025}
\textbf{Lagache} {et~al.} 2025, arXiv e-prints, arXiv:2506.15322.
\bibitem[{{Martig} {et~al.}(2009){Martig}, {Bournaud}, {Teyssier}, \&
  {Dekel}}]{Martig2009}
\textbf{Martig} {et~al.} 2009, \apj, 707, 250.
\bibitem[{{Oesch} {et~al.}(2023){Oesch}, {Brammer}, {Naidu}, {Bouwens},
  {Chisholm}, {Illingworth}, {Matthee}, {Nelson}, {Qin}, {Reddy}, {Shapley},
  {Shivaei}, {van Dokkum}, {Weibel}, {Whitaker}, {Wuyts}, {Covelo-Paz},
  {Endsley}, {Fudamoto}, {Giovinazzo}, {Herard-Demanche}, {Kerutt},
  {Kramarenko}, {Labbe}, {Leonova}, {Lin}, {Magee}, {Marchesini}, {Maseda},
  {Mason}, {Matharu}, {Meyer}, {Neufeld}, {Prieto Lyon}, {Schaerer}, {Sharma},
  {Shuntov}, {Smit}, {Stefanon}, {Wyithe}, \& {Xiao}}]{Oesch2023}
\textbf{Oesch} {et~al.} 2023, \mnras, 525, 2864.
\bibitem[{{Valentino} {et~al.}(2023){Valentino}, {Brammer}, {Gould}, {Kokorev},
  {Fujimoto}, {Jespersen}, {Vijayan}, {Weaver}, {Ito}, {Tanaka}, {Ilbert},
  {Magdis}, {Whitaker}, {Faisst}, {Gallazzi}, {Gillman}, {Gim{\'e}nez-Arteaga},
  {G{\'o}mez-Guijarro}, {Kubo}, {Heintz}, {Hirschmann}, {Oesch}, {Onodera},
  {Rizzo}, {Lee}, {Strait}, \& {Toft}}]{Valentino2023}
\textbf{Valentino} {et~al.} 2023, \apj, 947, 20.
\bibitem[{{Wang} {et~al.}(2019){Wang}, {Schreiber}, {Elbaz}, {Yoshimura},
  {Kohno}, {Shu}, {Yamaguchi}, {Pannella}, {Franco}, {Huang}, {Lim}, \&
  {Wang}}]{wangtao2019}
\textbf{Wang} {et~al.} 2019, \nat, 572, 211.
\bibitem[{{Weibel} {et~al.}(2024){Weibel}, {Oesch}, {Barrufet}, {Gottumukkala},
  {Ellis}, {Santini}, {Weaver}, {Allen}, {Bouwens}, {Bowler}, {Brammer},
  {Carnall}, {Cullen}, {Dayal}, {Dickinson}, {Donnan}, {Dunlop}, {Giavalisco},
  {Grogin}, {Illingworth}, {Koekemoer}, {Labbe}, {Marchesini}, {McLeod},
  {McLure}, {Naidu}, {P{\'e}rez-Gonz{\'a}lez}, {Shuntov}, {Stefanon}, {Toft},
  \& {Xiao}}]{Weibel2024}
\textbf{Weibel} {et~al.} 2024, \mnras, 533, 1808.
\bibitem[{{Xiao} {et~al.}(2024){Xiao}, {Oesch}, {Elbaz}, {Bing}, {Nelson},
  {Weibel}, {Illingworth}, {van Dokkum}, {Naidu}, {Daddi}, {Bouwens},
  {Matthee}, {Wuyts}, {Chisholm}, {Brammer}, {Dickinson}, {Magnelli}, {Leroy},
  {Schaerer}, {Herard-Demanche}, {Lim}, {Barrufet}, {Endsley}, {Fudamoto},
  {G{\'o}mez-Guijarro}, {Gottumukkala}, {Labb{\'e}}, {Magee}, {Marchesini},
  {Maseda}, {Qin}, {Reddy}, {Shapley}, {Shivaei}, {Shuntov}, {Stefanon},
  {Whitaker}, \& {Wyithe}}]{Xiao2024}
\textbf{Xiao} {et~al.} 2024, \nat, 635, 311.
\bibitem[{{Xiao} {et~al.}(2025){Xiao}, {Williams}, {Oesch}, {Elbaz},
  {Dessauges-Zavadsky}, {Marques-Chaves}, {Bing}, {Ji}, {Weibel}, {Bezanson},
  {Brammer}, {Casey}, {Cloonan}, {Daddi}, {Dayal}, {Faisst}, {Franx},
  {Glazebrook}, {Hutter}, {Kartaltepe}, {Labbe}, {Lagache}, {Lim}, {Magnelli},
  {Martinez}, {Maseda}, {Nanayakkara}, {Schaerer}, \&
  {Whitaker}}]{Xiao2025_spiral}
\textbf{Xiao} {et~al.} 2025, \aap, 696, A156.
\bibitem[{{Xiao} {et~al.}(2023){Xiao}, {Elbaz}, {G{\'o}mez-Guijarro}, {Leroy},
  {Bing}, {Daddi}, {Magnelli}, {Franco}, {Zhou}, {Dickinson}, {Wang},
  {Rujopakarn}, {Magdis}, {Treister}, {Inami}, {Demarco}, {Sargent}, {Shu},
  {Kartaltepe}, {Alexander}, {B{\'e}thermin}, {Bournaud}, {Ciesla}, {Ferguson},
  {Finkelstein}, {Giavalisco}, {Gu}, {Iono}, {Juneau}, {Lagache}, {Leiton},
  {Messias}, {Motohara}, {Mullaney}, {Nagar}, {Pannella}, {Papovich}, {Pope},
  {Schreiber}, \& {Silverman}}]{Xiao2023}
\textbf{Xiao} {et~al.} 2023, \aap, 672, A18.
\end{thebibliography}


\end{document}